\documentclass[a4paper,11pt]{article}
\usepackage{amssymb,amsmath,bm}
\usepackage{a4wide}
\usepackage{color}
\usepackage{slashed}
\usepackage{graphicx}
\usepackage{amsfonts}
\usepackage{lscape}
\usepackage{hyperref}
\usepackage{amsthm}
\hypersetup{
    colorlinks=true, 
    linktoc=all,     
    linkcolor=blue,  
    citecolor= blue
}
\usepackage{booktabs}
\usepackage{array}
\usepackage{rotating}
\usepackage[numbers, sort&compress]{natbib}
\usepackage{float}
\usepackage[utf8]{inputenc}
\usepackage[T1]{fontenc}
\usepackage{bm}% bold math

\usepackage[dvipsnames]{xcolor}

\newcommand{\beq}{\begin{equation}}
\newcommand{\eeq}{\end{equation}}
\newcommand{\bea}{\begin{eqnarray}}
\newcommand{\eea}{\end{eqnarray}}

\newcommand{\e}{{\rm e}}

\begin{document}

\begin{flushright}{\small IFIC/21-08,\quad FTUV-21-0407}
\end{flushright}

\vspace{1cm}

%=================
\begin{center}
	{\Large \bf \boldmath Implications of the Muon g-2 result on the\\[0.2cm] flavour structure of the lepton mass matrix} \\[1cm]
	{\large
	Lorenzo Calibbi$\,^\dag$\footnote{calibbi@nankai.edu.cn}, M.L.~L\'{o}pez-Ib\'{a}\~{n}ez$\,^{\ddag\, \S}$\footnote{maloi2@uv.es}, Aurora Melis$\,^\star$\footnote{aurora.melis@uv.es}, Oscar Vives$\,^\times $\footnote{oscar.vives@uv.es}
	\\[20pt]}
 	{\small 
 	$^\dag$ School of Physics, Nankai University, Tianjin 300071, China  \\[5pt]}
    {\small
    $^\ddag$ CAS Key Laboratory of Theoretical Physics, Institute of Theoretical Physics,\\Chinese Academy of Sciences, Beijing 100190, China \\[5pt]}
    {\small
    $^\S$ Departamento de Física, Campus de Rabanales Edif. C2,\\
    Universidad de Córdoba, E-14071 Córdoba, Spain \\[5pt]}
 	{\small
 	$^{\star}$ Laboratory of High Energy and Computational Physics, NICPB, R\"avala 10, 10143 Tallinn, Estonia  \\[5pt]}
 	{\small
 	$^{\times}$ Departament de F\'{i}sica T\`{e}orica, Universitat de Val\`{e}ncia,  Dr.~Moliner 50, E-46100 Burjassot\\ $\&$ IFIC, Universitat de Val\`{e}ncia $\&$ CSIC, E-46071,  Paterna, Spain  \\[5pt]}

\end{center}
\vspace*{1cm}
% =================

\begin{abstract}
\noindent
The confirmation of the discrepancy with the Standard Model predictions in the anomalous magnetic moment by the Muon g-2 experiment at Fermilab points to a low scale of new physics. Flavour symmetries broken at low energies can account for this discrepancy but these models are much more restricted, as they would also generate off-diagonal entries in the dipole moment matrix. Therefore, if we assume that the observed discrepancy in the muon $g-2$ is explained by the contributions of a low-energy flavor symmetry, lepton flavour violating processes can constrain the structure of the lepton mass matrices and therefore the flavour symmetries themselves predicting these structures.  We apply these ideas to several discrete flavour symmetries popular in the leptonic sector, such as $\Delta (27)$, $A_4$, and $A_5 \ltimes {\rm CP}$. 
\end{abstract}
\vspace*{0.5cm}

\newpage

% =================================================

\section{Introduction}
\setcounter{footnote}{0}

The new results from the Muon g-2 experiment at Fermilab~\cite{PhysRevLett.126.141801} have confirmed the long-standing discrepancy between the Standard Model (SM) prediction for the anomalous magnetic moment of the muon, $a_\mu = (g_\mu-2)/2$, and the previous BNL measurements \cite{Brown:2001mga,Aoyama:2020ynm}. This outcome makes it unlikely that the effect is due to a statistical fluctuation or overlooked systematics, hence making a strong case for new physics interacting with SM muons at scales not far above the electroweak (EW) scale.

The announcement of the very precise measure of the muon anomalous magnetic moment at BNL in 2001~\cite{Brown:2001mga}, already 2.6~$\sigma$ from the SM prediction started the search of new physics models to accommodate such a large discrepancy~\cite{Czarnecki:2001pv}.
The final result of BNL, published in 2006~\cite{Bennett:2006fi}, and the refinements of the SM prediction have led this discrepancy to 3.7~$\sigma$~\cite{Aoyama:2020ynm}:
\begin{align}
    \Delta a_\mu \equiv &~a_\mu^{\rm BNL} - a_\mu^{\rm SM} = (2.79 \pm 0.76) \times 10^{-9}. \label{eq:dam}
\end{align}

In fact, this discrepancy points naturally to a very low scale of new physics. This can be easily seen if we compare the observed discrepancy with the EW contribution in the SM,
\beq
a_\mu^{\rm EW} ({\rm 1 ~loop}) \simeq \frac{5}{3} \frac{G_F m_\mu^2}{8 \sqrt{2} \pi^2} \simeq 1.95 \times 10^{-9} \, .
\eeq
The size of this contribution, at one loop and with no special suppression factor, is determined by the mass of the $W$ boson, i.e.~80~GeV.

Indeed, the central value of the measured discrepancy in Eq.~(\ref{eq:dam}) is roughly a factor 1.5 larger. This implies that a new physics contribution with no special enhancement factor, couplings of electroweak size, and proportional to the muon mass would point to a mass scale of only 65 GeV!! It is true that some extensions of the SM have additional enhancement factors that allow a somewhat larger scale, for a recent review cf.~\cite{Lindner:2016bgg}. An example of this is supersymmetry, where the contributions to the anomalous magnetic moment are $\tan \beta$ enhanced which with some other ${\cal O} (1)$ factor may allow a mass scale $\lesssim 1$ TeV. As we show in section~\ref{subsec:DipoleStruct}, in the case of low-scale flavor symmetry, this scale, the flavor scale, is given by $\Lambda_f \simeq 515~\sqrt{\kappa}$~GeV with $\kappa$ a numerical factor that typically ranges from $1/8$ to $20$. Hence, the confirmation of this discrepancy by Fermilab implies that new physics coupling to the muon is expected not far above the TeV scale, unless large couplings give a chirally-enhanced contribution to the muon $g-2$, which
would require tuned cancellations between the tree-level contribution and a new radiative contribution to the muon mass~\cite{Capdevilla:2020qel,Capdevilla:2021rwo,Calibbi:2020emz}.

In this note, we show that, if confirmed, the Muon g-2 result would have profound implications, in the context of a low-energy flavor symmetry, for the flavour structure of the lepton sector of the Standard Model (SM), practically constraining the  charged-lepton Yukawa matrix, or the scale of flavour symmetry breaking.
The key point can be expressed as follows. 
Although so far we have no experimental indication of the mechanism responsible for the observed flavor structures, the previous success of symmetries in the advance of modern physics, makes us believe that the use of a flavor symmetry may help in the solution of the flavor puzzle. So, let us make the (to some extent strong) assumption that a
new flavour dynamics at the scale $\Lambda_f$ is at the origin of the hierarchies and peculiar patterns of fermion masses and mixing. As a specific example, we can have in mind a model based on flavour symmetries of the Froggatt-Nielsen type~\cite{Froggatt:1978nt,Leurer:1992wg,Leurer:1993gy}, where the SM Yukawa couplings arise after the symmetry is broken by the vev of scalar fields called flavons and such breaking is transmitted to the SM fermions through the interaction of certain mediator fields (cf.~\cite{Calibbi:2012yj,Calibbi:2012at} for further discussion about the UV completion of this kind of models).
Below this scale, we integrate out flavour mediators and flavons so that the only dimension four operator in the SM are the Yukawa couplings, and higher dimension operators are suppressed by the corresponding power of $\Lambda_f$.  Besides, the dipole operators $\sim \bar\ell \sigma_{\mu \nu} P_R \ell ~F^{\mu\nu}$ involve exactly the same fermionic fields as the Yukawa couplings, and, necessarily, the same flavour dynamics must dictate the flavour structure of these couplings are responsible for the $a_\mu$ anomaly.  If we call $\Lambda$ the scale associated with the new physics in the loop that generates the dipole operators, we have that provided that $\Lambda_f \gg \Lambda$, the flavour structure of the dipole operators will be exactly proportional to the Yukawa couplings (the only flavourful dimension 4 operator), with negligible corrections of order $m_\ell^2/\Lambda_f^2$. However, if $\Lambda_f \simeq \Lambda$, or even the new physics responsible for the new  contributions to dipole moments is the very same flavour dynamics, the flavour structure of the dipole operator will be similar to the Yukawa structure, but, in general, not exactly proportional to the Yukawa matrices. As a consequence, such a flavour dynamics (e.g.~a Froggatt-Nielsen-like model based on some flavour symmetries) will be severely constrained by the requirement that the muon dipole is as large as needed to account for the anomaly, simultaneously evading the stringent bounds on the flavour-violating counterparts of such operator, inducing the lepton flavour violating (LFV) decays $\ell\to\ell^\prime\gamma$ (for a recent review on LFV see~\cite{Calibbi:2017uvl}). In other words, the fact that the charged leptons Yukawas and the coefficients of the dipole operators need to be to large extent aligned to evade LFV constraints provides us with non-trivial information on any underlying flavour dynamics broken at a low scale.

The new result of the Muon g-2 experiment is 3.3$\sigma$ away from the SM prediction~\cite{PhysRevLett.126.141801} and thus is compatible with the previous BNL measurement, Eq.~(\ref{eq:dam}).
The combination of the two experiments reads: 
\begin{align}
    \Delta a_\mu \equiv &~a_\mu^{\rm NEW} - a_\mu^{\rm SM} = (2.51 \pm 0.59) \times 10^{-9}\,, \label{eq:damnew}
\end{align}
which amounts to a $4.2\sigma$ deviation from the SM prediction.
It is unlikely that such a discrepancy can be fully explained by underestimated uncertainties of the SM hadronic contribution~\cite{Aoyama:2020ynm}. It is also interesting to notice that, even if hadronic vacuum polarization effects were large enough to account for the anomaly, as suggested by recent lattice QCD results~\cite{Borsanyi:2020mff}, this would typically impact the EW fit causing tensions of similar significance in other EW observables~\cite{Passera:2008jk,Crivellin:2020zul,Keshavarzi:2020bfy,Malaescu:2020zuc}.

For what concerns the electron $g-2$ the situation is at present inconclusive. The SM calculation crucially depends on $\alpha_{\rm em}$, but the two most precise determinations of it from atomic physics~\cite{Parker:2018vye,Morel:2020dww} are incompatible at the $5\sigma$ level. Depending on which measurement is employed the prediction for $a_e$ gives:
\begin{align}
\Delta a_\e^{\rm Cs} \equiv &~a_\e^{\rm exp} - a_\e^{\rm SM,Cs} = -(8.8\pm 3.6) \times 10^{-13}, \label{eq:daeCs} \\
\Delta a_\e^{\rm Rb} \equiv &~a_\e^{\rm exp} - a_\e^{\rm SM,Rb} = (4.8\pm 3.0) \times 10^{-13}. \label{eq:daeRb} 
\end{align}
In particular, the second one~\cite{Morel:2020dww} is compatible with the experimental measurement at the $1.6~\sigma$ level and thus we will just employ it to bound possible new-physics contribution to $a_\e$.

As mentioned above, in an effective Lagrangian approach, non-standard effects to the leptonic observables of interest ($\Delta a_\ell$, $\mu\to \e\gamma$, EDMs, etc.) are captured by the dipole operators:
\beq
    {\cal L} \:\supset\: \frac{e\, v}{8\, \pi^2}\, C_{\ell \ell^\prime}\left(\bar\ell \sigma_{\mu \nu} P_R \ell^\prime\right)\, F^{\mu\nu} \:+\: {\rm h.c.} \quad\quad \ell,\ell^\prime =\e,\mu,\tau\,,
\label{eq:L-dipole}
\eeq
where $v\simeq 246~\text{GeV}$ is the Higgs vev.
Here we define the Wilson coefficients $C_{\ell \ell^\prime}$ (that in our convention have mass dimension $\rm GeV^{-2}$) as corresponding to new-physics contributions only.
This effective Lagrangian constitutes a model-independent description of the new-physics effects we are interested in, so long as the new-physics scale is much larger than the energy scale associated to our observables, i.e.~the lepton masses. In terms of the above Wilson coefficients the new-physics contribution to the $\Delta a_\ell$ reads:\footnote{This is obtained employing the conventions found e.g.~in~\cite{Aebischer:2021uvt}. Notice the sign difference (stemming from the definition of the covariant derivative) with respect to the result reported in the review in Ref.~\cite{Jegerlehner:2009ry}.}
\beq
    \Delta a_\ell = \frac{{{m_{\ell}}v}}{2 \pi^2}\, \mbox{\rm Re} (C_{\ell \ell}).
\label{eq:AC}
\eeq
In order to fit the experimental result in~Eq.(\ref{eq:damnew}), the  relevant dipole coefficient needs to attain the following 1$\sigma$ range:
 \bea
    \mbox{\rm Re}(C_{\mu\mu})   ~ \approx ~ [1.5,\,2.4] \times 10^{-9}~ ~\mbox{\rm GeV}^{-2}. \label{eq:Cmm}
\eea
For what concerns $a_e$, the result in Eq.~(\ref{eq:daeRb}) translates into the 95\% CL bound:
\bea
 -1.9\times  10^{-11}~ ~\mbox{\rm GeV}^{-2}~\lesssim ~   \mbox{\rm Re}(C_{ee})   ~ \lesssim ~ 1.7 \times 10^{-10}~ ~\mbox{\rm GeV}^{-2}. \label{eq:Cee}
\eea
Finally the yet unmeasured tau $g-2$ provide a loose bound to the $C_{\tau\tau}$ coefficient:
\bea
 -3.7\times 10^{-4} ~\mbox{\rm GeV}^{-2}~\lesssim ~   \mbox{\rm Re}(C_{\tau\tau})   ~ \lesssim ~1.7\times 10^{-4} ~\mbox{\rm GeV}^{-2}, \label{eq:Ctt}
\eea
which follows from the constraint $-0.007 <a_\tau < 0.005$
that has been obtained employing data for tau lepton production at leptonic colliders~\cite{GonzalezSprinberg:2000mk} and the SM prediction $a^{\rm SM}_\tau = 117721(5)\times 10^{-8}$~\cite{Eidelman:2007sb}.

On the other hand, the imaginary parts of the dipole coefficients give rise to lepton EDMs:
\beq
    d_\ell = \frac{e v}{4 \pi^2}\, \mbox{\rm Im} (C_{\ell \ell})\,,
\eeq
resulting in the following bounds
\beq
|\mbox{\rm Im} (C_{ee})|\lesssim 9.0\times 10^{-17}~ \mbox{\rm GeV}^{-2},~
|\mbox{\rm Im} (C_{\mu\mu})|\lesssim 1.2\times 10^{-6}~ \mbox{\rm GeV}^{-2},~
|\mbox{\rm Im} (C_{\tau\tau})|\lesssim 3.7\times 10^{-4}~ \mbox{\rm GeV}^{-2},
\eeq
which were obtained respectively from the experimental limits $d_e <1.1\times 10^{-29}\,e\,\text{cm}$~\cite{Andreev:2018ayy}, $d_\mu <1.9\times 10^{-19}\,e\,\text{cm}$~\cite{Bennett:2008dy},
$d_\tau <4.5\times 10^{-17}\,e\,\text{cm}$~\cite{Inami:2002ah}.
As we can see, the very stringent constraint on the electron EDM implies that, if $C_{ee}$ saturates the bound in Eq.~\eqref{eq:Cee}, its phase must be suppressed at the $10^{-6}-10^{-5}$ level.\footnote{Note that Ref.~\cite{Alonso-Gonzalez:2021jsa}, applying a similar strategy as the one adopted here, assumed that a flavour symmetry dictates the flavour structure of the dipole operators, in order to constrain their imaginary parts,~i.e.~the EDMs.}

The flavour-changing couplings instead contribute to LFV processes, in particular to the radiative decays:
\begin{align}
\frac{{\rm BR}(\ell\to {\ell^\prime} \gamma)}{{\rm BR}(\ell\to \ell^\prime \nu \bar{\nu}^\prime)} = \frac{3\alpha}{\sqrt{2}\pi \,G_F^3\, m_\ell^2} \left( 
|C_{\ell \ell^\prime}|^2 + |C_{\ell^\prime \ell}|^2\right)\,,
\end{align}
where the coefficients $C_{\ell \ell^\prime}$ are defined in the basis where the lepton Yukawa matrix $Y_\ell$ is diagonal, that we call below the ``mass basis''.
The experimental bound ${\rm BR}(\mu\to \e \gamma) < 4.2\times 10^{-13}$ \cite{TheMEG:2016wtm} then translates into the following constraint:
\bea \label{eq:Cem}
    |C_{\e\mu}|,~| C_{\mu\e}| &\lesssim &  3.9\times 10^{-14}~~ \mbox{\rm GeV}^{-2}\,.
\eea
Similarly, bounds from $\tau\to\mu\gamma$ and $\tau\to e\gamma$~\cite{Abdesselam:2021cpu,Aubert:2009ag} respectively give:
\bea \label{eq:Cmt}
    |C_{\tau\mu}|,~| C_{\mu\tau}| & \lesssim & 5.0\times 10^{-10}~~ \mbox{\rm GeV}^{-2}\,, \\
        |C_{\tau e}|,~| C_{\mu e}| & \lesssim  & 4.3\times 10^{-10}~~ \mbox{\rm GeV}^{-2}\,.
        \label{eq:Cet}
\eea

As mentioned above, all these constraints refer to coefficients in the lepton mass basis, namely the basis where the charged lepton Yukawa matrix is diagonal, which is connected to any other basis by unitary rotations of the lepton fields:
\bea
Y_\ell^\text{diag} = V^\dag Y_\ell W = \text{diag}(y_e,y_\mu,y_\tau)\,,
\label{eq:biunit}
\eea
where $V$ and $W$ are the left- and right-handed rotations, respectively. 
This follows from our convention for the Yukawa interactions of the leptons with the Higgs: $(Y_\ell)_{ij}\, \bar{L}_i E_j \Phi$, where $L_i$ ($E_j$) are the $SU(2)_L$ doublet (singlet) lepton fields.
In particular, we consider here the biunitary transformation 
as in Eq.~\eqref{eq:biunit} connecting the mass basis and the flavour basis, {\it i.e.} the basis defined by the symmetry invariant interactions of the flavons (and their vacuum expectation values) with the SM fermions and heavy mediators at the origin of the Yukawa interactions. Denoting  as $\widetilde C$ the dipole coefficients in 
the flavour basis (whose structure is dictated by the flavour symmetry) the same transformation to the mass basis gives the matrix of Eq.~\eqref{eq:L-dipole}:
\bea
C_{\ell\ell^\prime} = \left(V^\dag \widetilde C W\right)_{\ell\ell^\prime}\,,\quad\quad \ell,\ell^\prime =\e,\mu,\tau\,.
\eea
It is the elements of this matrix that are constrained
as shown in Eqs.~(\ref{eq:Cmm}-\ref{eq:Ctt}) and~(\ref{eq:Cem}-\ref{eq:Cet}).

%=================
\section{Charged lepton Yukawa structure}
\label{subsec:ChargStruct}
%=================

The only restriction we have on the charged lepton Yukawa matrix comes from the observed eigenvalues and, in principle, from the neutrino mixing. Nevertheless, it is always possible to generate the neutrino mixing entirely from the Majorana neutrino effective mass matrix or the neutrino Yukawa in case the neutrino masses are Dirac-like, and then,  we can not use these large mixing to constrain the entries in the charged-lepton matrix. Still, the strategy typically used for quark Yukawas in flavour symmetry models is to require that the smallness of CKM mixing is not due to a precise cancellation between large mixing angles in the flavor basis both for the up and down Yukawa matrices, but to the smallness of the mixings in both matrices in this flavor basis. This strategy is not useful here with the neutrino mixing angles, as neutrino mixings are ${\cal O}(1)$ with the partial exception of $\theta_{13}$. However, we can still follow a similar strategy with the charged-lepton eigenvalues: the hierarchy in the eigenvalues is due to a corresponding hierarchy in the Yukawa matrix elements in the flavour basis. This means that we require $m_\mu/m_\tau \simeq 0.059$ or $m_e/m_\mu \simeq 0.005$ to be due to different contributions ${\cal O}(0.05)$ or ${\cal O}(0.005)$ from the elements of the Yukawa matrix and not to a finely-tuned cancellation of several ${\cal O}(1)$ contributions, a requirement that has been formally formulated in~\cite{Marzocca:2014tga}. 

If we take the Cabibbo angle as our expansion parameter $\lambda = 0.2$,\footnote{Clearly, the physics is completely independent of the choice of this expansion parameter. We choose the Cabibbo angle aiming at a model valid for both quarks and leptons, along the lines of several examples already present in the literature \cite{Ross:2004qn,deMedeirosVarzielas:2006fc,deMedeirosVarzielas:2018vab,Cooper:2012wf,Feruglio:2007uu, CarcamoHernandez:2021osw}.} suppressing ${\cal O}(1)$  coefficients and normalising all the matrix with the $(3,3)$ element ($\approx y_\tau$), we have,   
\beq
    Y_\ell/y_{33} ~=~ \begin{pmatrix}
              \lambda^a& \lambda^b&\lambda^c\\  \lambda^d& \lambda^e&\lambda^f\\  \lambda^g& \lambda^h& 1
            \end{pmatrix}, 
\eeq
where, we can always order rows and columns such that the larger $2\times 2$ sub-determinant is in the 2--3 sector. Then, from $m_\mu/m_\tau = 0.059$, we can require that the determinant of the 2--3 submatrix is $\lambda^e -\lambda^{f+h} = \lambda^2$. Barring accidental cancellations, this implies, $e \geq 2$ and $f+h \geq 2$, where at least one of these inequalities must be turned into equality. Notice that the condition of larger 2--3 subdeteminant  implies that we can also apply these conditions exchanging the first two rows or columns, and then we have additionally, $b\geq 2$, $d\geq 2$, $h + c \geq 2$ and $f+g\geq 2$.

Then, in either case, $e=2$ or $f+h=2$, the requirement of determinant of the $3 \times 3$ matrix to be $\lambda^7$ (from $m_e m_\mu/m_\tau^2 \simeq \lambda^7$), implies that $a \geq 5$. For the same reason $b+d \geq 7$, with both $b\geq 2$ and $d\geq 2$. Then, redefining the exponents, we have,
\beq
    Y_\ell/y_{33} ~=~ \begin{pmatrix}
              \lambda^{5+a}& \lambda^{2+b}&\lambda^c\\  \lambda^{2+d}& \lambda^{2+e}&\lambda^f\\  \lambda^g& \lambda^h& 1
            \end{pmatrix}, 
\label{eq:Ye}
\eeq
with $a,b,c,d,e,f,g,h \geq 0$.

Additional conditions from the $2\times 2$ and $3\times3$ determinants, are $f+h \geq2$, $f+g \geq2$, $h+c\geq 2$,  $b+ d \geq 3$, $c+g+e\geq 5$, $f+b+g \geq 5$ and $h+d+c \geq 5$. In principle, this would be all the information we have on the charged-lepton Yukawa from the requirement of hierarchical masses in the flavor basis. However, in most of the cases, we expect that the (3,3) entry is the only $O(1)$ element, {\it i.e.} $c,f,g,h > 0$. In the case of non-Abelian symmetries, where flavons belong to dimension 2 or dimension 3 representation, we can always choose the basis where the largest vev sits in the (3,3) position of the Yukawa matrix. A possible exception could be Abelian symmetries where the three generations of left-handed or right-handed fields have the same charge, but in this case, the three rows or columns would be of the same order in $\lambda$ (notice that with different $O(1)$ coefficients you could still have correct charged-lepton masses). With the exception of this special case, we will consider in the following hierarchical matrices with $c,f,g,h \geq 1 $.

%=========================================================================
\section{Flavour  structure of the dipole matrix}
\label{subsec:DipoleStruct}
%=========================================================================
As we discussed in the introduction, a new contribution to the anomalous magnetic moment from a low-energy flavour symmetry is possible and it will have a flavour dependence similar to the SM Yukawa couplings but with different $\mathcal{O}(1)$ coefficients in the corresponding operators, with or without a photon, with the same flavon fields. Here, we explain the intricate structure of Yukawa couplings through a Froggatt-Nielsen (FN) symmetry. In this framework, it looks completely natural to use the same mechanism to explain the structure of the dipole operators. From the point of view of the flavour symmetry, Yukawas and dipole moments involve exactly the same SM fields with the obvious exception of the photon. Then, it is clear that the flavour ``charges'' to be compensated by flavon fields are exactly the same, and we can expect the same flavon fields to enter in the two operators. 

In fact, fermion masses and anomalous magnetic moments are intimately connected.
Any radiative correction to the fermion masses gives a contribution to the anomalous magnetic moment if we attach a photon to one of the internal lines. 
Although the FN contributions to the Yukawa matrices usually considered are tree-level diagrams, loop corrections to the tree-level diagrams are always present, and under certain conditions they can be sizeable with respect to the tree-level processes. 
For instance, this is the case in flavour models where some larger flavon VEVs appear in the loop diagrams but not in the FN transitions \cite{Calibbi:2020emz}, as we discuss in section~\ref{sec:toy}.
%Now, it is clear that this loop diagram generating a loop correction to the Yukawa would be the same as the diagram generating the dipole coefficients simply adding a photon.
In general, we expect a contribution to the anomalous magnetic moment $\Delta a_\ell = C m_\ell^2/M^2$~\cite{Czarnecki:2001pv}, with $C$ a loop factor of order $1/(16 \pi^2)$, if the fermion mass is mostly produced by the tree-level diagram%if the fermion mass is present at tree-level
, or $C\sim {\cal O}(1)$, %if the mass is generated at loop level~\cite{Okada:2013iba}. 
if the tree-level diagram is absent or when the loop diagrams compete with it \cite{Okada:2013iba}. 
In our flavour symmetry models, we could have radiative corrections to the mass similar to the tree-level contribution which implies that a large contribution to $a_\ell$, with $C\sim {\cal O}(1)$, can be expected.\footnote{If we allow the tree level Yukawa and loop contribution of opposite signs such that the observed mass is due to a certain cancellation between both contributions, we could allow for a somewhat larger $C$~\cite{Calibbi:2020emz}.} 
Still, it is evident that this correspondence between masses and dipole moments is not restricted to the diagonal Yukawa matrices in the mass basis, but can be equally applied to the whole Yukawa matrix. 

Our strategy, in the scenario of a flavor symmetry described above, is then the following: 
\begin{enumerate}
\item We assume that the 4.2~$\sigma$ discrepancy in the muon anomalous magnetic moment is due to a low-energy flavour symmetry (or at least the flavour structure of the new physics responsible for it is controlled by the flavour symmetry at the origin of the Yukawa couplings). 
\item The breaking of the flavour symmetry generates the observed charged-lepton Yukawa matrices in the flavour basis, and, simultaneously the dipole matrix in this basis.
\item As we have explained, we can expect the flavons entering both matrices to be the same, given they are identical from the symmetry point of view, but we expect in general different ${\mathcal O}(1)$ coefficients in the operators with identical flavon fields with or without a photon coupling, contributing respectvely to the dipole or Yukawa matrices.\footnote{These ${\mathcal O}(1)$ coefficients are due to the different ways to couple the the photon inside the loop or the different ways to close the loop, as explicitly shown in \cite{Calibbi:2020emz}.}

\item Rotations to the mass basis will digonalise the Yukawa couplings, but the dipole matrix will not be diagonalised. 
\end{enumerate}
If these conditions are realised, the anomalous magnetic moments provide a hint on the scale of flavour symmetry breaking and off-diagonal LFV dipole operators will provide a strict bound on the flavour structure of the charged-lepton Yukawa matrix.

At this point, we must remind the reader that, even though it is customary to choose the third generation to sit in the $(3,3)$ element of the Yukawa matrix, we could have chosen any other basis by using arbitrary combinations of fermion fields through unitary rotation. As it is well-known, physical observables are always independent of the choice of basis. These unitary rotations
between fermion fields affect in the same way Yukawa and dipole matrices and, then, they cancel exactly when diagonalizing the Yukawa matrix to calculate dipole observables. For a complete discussion and a fully basis independent formulation see, for instance, Ref.~\cite{Botella:2004ks}.

In fact, as we will see below, in most of the phenomenologically realistic cases (with a low-scale flavor symmetry explaining the $(g-2)_\mu$ discrepancy) the Yukawa matrices are hierarchical and approximately diagonal. Then, in most of these cases, the rotation to the mass basis will only modify $\mathcal{O}(1)$ coefficients in the off-diagonal entries. 
There are situations, however, where a specially small entry in the dipole matrix receives larger contributions from other elements through this rotation. 
Nevertheless, although it is always necessary to rotate the matrix to the mass basis, it is easy to see that, due to the different $\mathcal{O}(1)$ elements with respect to the Yukawa matrices, the size of the entry in the mass basis will be of the same order or larger than the entry in the flavor basis (barring accidental cancellations).

Now, following the previous discussion, we can assume that  the entries of the dipole matrix are larger or of the order of the corresponding entries of the Yukawa matrix in the flavor basis,
\beq
    C_{\ell \ell^\prime} ~\gtrsim~ \kappa ~ \frac {y_\tau}{\Lambda_f^2} \begin{pmatrix}
              \lambda^{5+a}& \lambda^{2+b}&\lambda^c\\  \lambda^{2+d}& \lambda^{2+e}&\lambda^f\\  \lambda^g& \lambda^h& 1
            \end{pmatrix}, 
    \label{eq:CD}        
\eeq
where we omitted different ${\cal O} (1)$ coefficients that, in general, appear in the different entries.
Here $\kappa$ is a global factor that takes care of the relative size of the tree-level Yukawa (or the observed mass) and the loop contribution to the mass. Using Eq.~(\ref{eq:AC}), we have $C \simeq \kappa /(2 \pi^2)$. Then, $\kappa \simeq 1/8$ if the dipole is loop suppressed with respect to the mass, but $\kappa \simeq 2 \pi^2 \simeq 20 $ if the mass has a radiative origin. 

As shown in the introduction, experimental measurements are directly constraining the above dipole matrix, and thus also the flavour structure 
of the Yukawa matrix in Eq.~\eqref{eq:Ye}, in the flavor basis. In particular the $C_{\mu\mu}$ entry can be set by requiring an effect sufficiently large to account for the muon $g-2$ anomaly, as in Eq.~\eqref{eq:Cmm}, while off-diagonal entries are bounded by the LFV processes, as shown in~Eqs.~(\ref{eq:Cem}-\ref{eq:Cet}).
As will be clear from the following discussion, the latter bounds affect the off-diagonal entries of the two matrices to such an extent that, in this scenario, the dominant contributions 
to muon and electron masses can only arise from the diagonal entries of $Y_\ell$. Hence, in order to avoid $y_e$ and $y_\mu$ to be suppressed below the observed values, we must take $a=e=0$.
This simplifies the discussion: first, imposing
a solution of the muon $g-2$ anomaly at the $1\sigma$ level gives $C_{\mu\mu}\simeq \kappa~ y_\tau \lambda^{2} / \Lambda_f^2 \approx 1.5\times 10^{-9}/\text{GeV}^2$,
where $\lambda= 0.2$ and $y_\tau = \sqrt{2} m_\tau/v$. Hence
the scale of the mediators is set to $\Lambda_f \approx 515 ~\sqrt{\kappa}$~GeV, where it is very important to emphasize once more that the $(g-2)_\mu$ discrepancy is reproduced with a larger flavour scale if the mass has a radiative origin, $\kappa \simeq 20$.
Then, besides unknown $\mathcal{O}(1)$ coefficients, this fixes the relative size of all the other entries of our matrix in terms of few parameters (the powers of $\lambda$) that we can now constrain. In particular, we can estimate the exponents $b,d$ from the requirements in Eq.~\eqref{eq:Cem}, which translate to  $C_{e\mu}/C_{\mu\mu} \simeq \lambda^b \lesssim 2.6 \times 10^{-5}$ and $C_{\mu e}/C_{\mu\mu} \simeq \lambda^d \lesssim 2.6 \times 10^{-5}$, that is $d, b \gtrsim 6.6$. Similarly we can employ the limits on the other dipole operators and obtain the following constraints on the flavour structure of the lepton Yukawa matrix:

\beq
    Y_\ell\approx y_\tau
    \begin{pmatrix}
              \lambda^{5}& \lesssim\lambda^{8.6}& \lesssim\lambda^{2.8} \\   \lesssim\lambda^{8.6} & \lambda^{2}&\lesssim\lambda^{2.7}\\ \lesssim\lambda^{2.8}& \lesssim\lambda^{2.7}& 1
            \end{pmatrix}, 
    \label{eq:Ye-bounds}        
\eeq
where again we set $\lambda = 0.2$. 
 The limits of Eq.~(\ref{eq:Ye-bounds}) apply directly on the exponents of Eq.~(\ref{eq:Ye}) 
and, more generally, on every product of the off-diagonal elements that could be introduced by the mass-basis rotation. 
Therefore we have $(b,d) \gtrsim 6.6$, $(c,g)\gtrsim 2.8$, $(f,h)\gtrsim 2.7$ and $(a,e)=0 $, additional constraints are $c+h\gtrsim 8.6$ and $f+g\gtrsim 8.6$.
As we can see from this result, only flavour models inducing a Yukawa matrix to large extent diagonal in the flavour basis (in particular in the 1-2 sector) can be compatible with the observed $g-2$ discrepancy. For instance, we can see from here that the ``original'' Froggatt-Nielsen $U(1)$ symmetry (that, for the sake of the observed neutrino mixing, requires $\mathcal{O}(1)$ rotations in the left-handed sector of this matrix~\cite{Altarelli:2012ia}) could never provide a texture fulfilling the above bounds.

Besides constraining the flavour structure of the Yukawa matrix, once the muon $g-2$ sets the entry $C_{\mu\mu}$ of  Eq.~\eqref{eq:CD}, we can also predict (still up to unknown $\mathcal{O}(1)$ coefficients) the entries $C_{ee}$ and $C_{\tau\tau}$, that is, the electron and tau $g-2$:
\begin{align}
    C_{ee} & ~\approx~ C_{\mu\mu}\, \lambda^3~ = ~1.2\times 10^{-11}~\mbox{\rm GeV}^{-2} ~~\Rightarrow ~~\Delta a_e \approx 7.8\times 10^{-14}\\
    C_{\tau\tau} & ~\approx~ C_{\mu\mu}/\lambda^2 ~=~ 3.9\times 10^{-8}~\mbox{\rm GeV}^{-2} ~~\Rightarrow~~ \Delta a_\tau \approx 8.5\times 10^{-7}\,.
\end{align}
In particular, the electron $g-2$ is less than an order of magnitude below the present sensitivity.
Notice that, given the structure of the matrix $C_{\ell\ell^\prime}$ dictated by the flavour symmetries, such predictions respect the ``naive scaling'' $\Delta a_\ell /\Delta a_{\ell^\prime} \sim (m_\ell/m_{\ell^\prime})^2$ discussed in~\cite{Giudice:2012ms}.\footnote{In the context of low-energy flavour models, this pattern can be however modified by the interplay between tree-level and radiative contributions to lepton masses. This can for instance enhance $\Delta a_e$ above the naive scaling and even flip its sign, as shown in~\cite{Calibbi:2020emz}.}

%=================
\section{Flavour symmetries}
\label{subsec:flavsyn}
%=================
In this section, we will analyse several popular flavour symmetries in neutrino physics (for reviews cf.~\cite{Altarelli:2010gt,Feruglio:2021sir}) to see if they can be responsible of the observed discrepancy in the anomalous magnetic moment of the muon while, at the same time satisfying the bounds on LFV processes
that we translated in the constraints shown in
Eq.~\eqref{eq:Ye-bounds}.

Naturally, the flavour symmetry contribution to the anomalous magnetic moment may not account for the whole discrepancy, as minimal flavour violating contributions, exactly proportional to the charged-lepton Yukawas are always possible. %although they would not be able to give a simultaneous explanation of electron and muon discrepancies. 
In this case, we can make the flavour symmetry contribution to the anomalous magnetic moment, and consequently to the LFV processes arbitrarily small, with a large enough breaking scale. 

%=================
\subsection{$\Delta (27)$ symmetry}
\label{subsec:D27}
%=================
Models based on a discrete $\Delta(27)$~\cite{deMedeirosVarzielas:2006fc,Ma:2006ip,Luhn:2007uq,Bjorkeroth:2015uou, deMedeirosVarzielas:2017sdv,deMedeirosVarzielas:2018vab,CarcamoHernandez:2018djj} have been very popular as an explanation of the observed large mixing in the lepton sector, while at the same time providing a simultaneous explanation of quark mixing and being compatible with grand unification. Their main feature is the natural mechanism to obtain the appropriate flavon alignment of the triplet fields~\cite{Bjorkeroth:2015uou},
\bea
    \langle\phi_{3}\rangle^T \,=\, v_3\, (0,\, 0,\, 1),\qquad
    \langle\phi_{23}\rangle^T \,=\, \frac{v_{23}}{\sqrt{2}}\, (0,\, 1,\, 1),\qquad \langle\phi_{123}\rangle^T \,=\, \frac{v_{123}}{\sqrt{3}}\, (1,\, 1,\, -1).
\eea
In this way, we obtain the charged-lepton mass matrices in the flavor basis defined by these vevs, as
\beq
    % M_e ~=~ v\; \left[\, 
    %         y_1^e\, \begin{pmatrix}
    %                     0 & 1 & 1 \\
    %                     1 & 2 & 0 \\
    %                     1 & 0 & -2
    %                 \end{pmatrix} \;+\;
    %         y_2^e\, \right]
  %  M_e ~=~ \frac{v}{\sqrt{2}}\; \left[\, 
    Y_\ell ~=~ \left[\, 
            y_1^e\, \begin{pmatrix}
                        0 & 0 & 0 \\
                        0 & 1 & 1 \\
                        0 & 1 & 1
                    \end{pmatrix} \;+\;
            y_2^e\, e^{i\eta}\,  \begin{pmatrix}
                        1 & 3 & 1 \\
                        3 & 9 & 3 \\
                        1 & 3 & 1
                    \end{pmatrix} \;+\;
            y_3^e\, e^{i\eta'}\,  \begin{pmatrix}
                        0 & 0 & 0 \\
                        0 & 0 & 0 \\
                        0 & 0 & 1
                    \end{pmatrix}  \,\right]    
\eeq
%Eq. (5.1c) in Bjork
with $y^e_3 \simeq 4 \times 10^{-2}$,   $y^e_2 \simeq 1 \times 10^{-5}$ and  $y^e_1 \simeq 2 \times 10^{-3}$ \cite{Bjorkeroth:2015uou}.

As we explained in the previous section, if a low-scale flavour symmetry is responsible for the observed discrepancy with the SM predictions in the muon anomalous magnetic moment, the dipole matrix has the same structure in the flavour basis with different ${\cal O} (1)$ coefficients. Therefore, in this case of small rotations, we can see that the structure of the dipole matrix remains unchanged in the mass basis and we have,
\beq
    C_{\ell \ell^\prime} ~=~ \kappa ~\frac {y^e_3}{\Lambda_f^2} \begin{pmatrix}
              \beta& 3 \beta&\beta\\ 3 \beta& \alpha + 9 \beta&\alpha + 3 \beta \\ \beta& \alpha+ 3 \beta& 1
            \end{pmatrix}, 
    \label{eq:Ce}        
\eeq
with $\alpha = y^e_1/y^e_3 = 0.05$ and $\beta=y^e_2/y^e_3 = 0.00025$.
In this case, we have $C_{e\mu}/C_{\mu\mu} = 0.015 \gg 2.6 \times 10^{-5}$, and therefore the model does not satisfy the constraint from $\mu \to e \gamma$. 

Indeed, this could be expected, as the $\Delta (27)$ symmetry generates off-diagonal entries in the charged-lepton mass matrix. Even taking into account that these off-diagonal entries are relatively small, the very stringent  constraints from $\mu \to e \gamma$ require, in practice, diagonal charged-lepton mass matrix in flavour symmetries broken at low energies,
as we can see from Eq.~\eqref{eq:Ye-bounds}.

%=================
\subsection{$A_4$ symmetry}
\label{subsec:A4}
%=================

The observation of neutrino mixing angles being close to the so called tri-bimaximal mixing has made the use of discrete symmetries to explain them very popular. Among them, $A_4$ has been very successful \cite{Babu:2002dz,Altarelli:2005yx,Adhikary:2006wi,Hirsch:2007kh,Altarelli:2008bg,Lin:2008aj,Honda:2008rs,Grimus:2008tm,Altarelli:2009kr,Morisi:2009qa,Antusch:2010es,Morisi:2013eca}. In this model, the $A_4$ symmetry is broken by different flavons for the neutrino and charged-lepton matrices, such that, at leading order, these sectors are separately invariant under two different residual symmetries, $G_e$ and $G_\nu$ and this fact can help to satisfy the requirement of nearly diagonal charged-lepton matrices.

The group $A_4$ can be generated by two elements, $T$ and $S$~\cite{Altarelli:2009kr}. Then, following this Ref.~\cite{Altarelli:2009kr}, charged-lepton masses are diagonal, at leading order in the basis where the $T$ generator is diagonal, while the neutrino masses present tri-bimaximal mixing if the flavons entering the neutrino matrices are invariant under the $S$ generator of $A_4$. In this way, the leptonic Yukawa coupling is diagonal and hierarchical at leading order. The hierarchies among generations are given by powers in $v_T/\Lambda_f=\epsilon\sim \lambda^2$, with $v_T$ the vev of the flavons invariant under $T$. However, as we have seen in Eq.~\eqref{eq:Ye-bounds}, LFV constraints require it to be diagonal up to order $\approx\lambda^6$ and therefore sub-leading terms must be considered. 

Off-diagonal elements in the Yukawa matrix in the flavor basis are suppressed by $v_S/\Lambda_f = \epsilon^\prime \lesssim \lambda^2$, where $v_S$ is the vev of the flavon invariant under $S$ that generates the tri-bimaximal mixing for neutrinos. 
The charged-lepton Yukawa matrix, again neglecting ${\cal O} (1)$ coefficients is then,
\beq
    Y_\ell ~=~ y_{33} \begin{pmatrix}
            \epsilon^2 & \epsilon \, \epsilon^\prime & \epsilon^\prime \\ \epsilon^2  \epsilon^\prime & \epsilon & \epsilon^\prime \\   \epsilon^2  \epsilon^\prime&\epsilon \, \epsilon^\prime & 1
            \end{pmatrix}, 
\label{eq:YeA4}
\eeq
with $\epsilon \simeq 0.05$ and $\epsilon^\prime \simeq 0.01$--0.05 \cite{Altarelli:2009kr,Feruglio:2008ht}. Therefore, in Eqs.~\eqref{eq:Ye} and \eqref{eq:CD}, we would obtain $b\simeq 3$ and $d\simeq 5$, still too large if we take into account the constraints from LFV. Notice here, that again the rotation to the mass basis would not change the size of the offdiagonal elements and no additional restrictions appear.

Unfortunately these constraints mean that this ``sequestering'' mechanism, which generates neutrino mixing close to tri-bimaximal mixing, would not be enough in typical models to suppress LFV below the observed limits in our scenarios.

%=================
\subsection{\boldmath $A_5\ltimes {\rm CP}$ symmetry}
\label{subsec:A5CP}
%=================
A variation of the models in section~\ref{subsec:A4} introduces a generalised CP symmetry as part of the flavour group \cite{Feruglio:2012cw,Holthausen:2012dk}; for instance, let us consider $A_5 \ltimes {\rm CP}$ like in Ref.~\cite{DiIura:2018fnk} (see also \cite{Ballett:2015wia, DiIura:2015kfa, Li:2015jxa}).

In this case, the breaking of the total symmetry follows similar lines to the previous $A_4$ model, where charged leptons and neutrinos remain invariant under the transformations of two different residual groups $G_e$ and $G_\nu$.
Those remnant subgroups of $A_5$ and CP fully determine the mixing matrix in each sector, up to permutations of rows and columns.
In particular, the formulations analysed in~\cite{DiIura:2018fnk} consider the solution with $G_e=Z_5$ and $G_\nu=Z_2\times {\rm CP}$, corresponding to the CP-violating scenario that better fits the experimental PMNS matrix in~\cite{DiIura:2015kfa}.
This choice implies a diagonal Yukawa matrix for charged leptons at leading order in the flavor basis defined by our $A_5\ltimes {\rm CP}$ symmetry,
\beq \label{eq:YeA5CP}
    Y_\ell ~=~ y_\tau \begin{pmatrix}
                        \lambda^5 &      0    & 0 \\
                         0    & \lambda^2 & 0 \\
                         0    &      0    & 1
                   \end{pmatrix},
\eeq
while the mixing pattern of neutrinos departs from the pure Golden Ratio mixing, characteristic of $A_5$, by a rotation in the 1-3 sector related to the CP group.
Adjusting the size of this rotation, the reactor angle can be successfully reproduced.

Two separate set of flavons for charged leptons, $\phi_e^i$, and neutrinos, $\phi_\nu^i$, are needed to realise this type of models.
The structure of their vevs is determined by the residual symmetries \cite{DiIura:2018fnk}.
At leading order, each set couples exclusively to the corresponding lepton fields (charged or neutral).
Next-to-leading order operators, however, generally include interactions between the {\it wrong} flavons and the other sector.
Corrections coming from those operators can spoil the desired mixing.

A possible way to keep them under control has been discussed in \cite{Lopez-Ibanez:2019rgb} and consist of introducing an additional $Z_N$ symmetry.
By choosing a suitable combination of charges for the fields, the NLO contribution to the charged-lepton masses is schematically given by
\beq \label{eq:dMe_A5CP}
    \delta Y_\ell ~=~ y_\tau\, \varepsilon_\nu^{n}\: \begin{pmatrix}
                                1 & 1 & 1 \\
                                1 & 1 & 1 \\
                                1 & 1 & 1
                            \end{pmatrix},
\eeq
where $\varepsilon_\nu=\langle \phi_\nu \rangle/\Lambda_f$ and the $1-$entry matrix symbolises a matrix whose elements are ${\cal O}(1)$ but not necessarily equal.
The size of $\varepsilon_\nu$ depends on the specific mechanism generating the light neutrino masses, but must always be within the usual perturbative range $\varepsilon_\nu \lesssim 0.3$ \cite{Lopez-Ibanez:2019rgb}.
Instead, the $n$ exponent can be made as large as needed (by an adequate charge assignment under $Z_N$) so that the off-diagonal entries in the dipole matrix of Eq.~\eqref{eq:CD}, $\left(b, c, d, f, g, h\right)$, can be conveniently arranged to avoid the LFV constraints displayed in Eq.~\eqref{eq:Ye-bounds} at the required level. Once more, the rotation to the mass basis has no effect on the size of the offdiagonal elements for realistic values of $n$.

For instance, if $\varepsilon_\nu\simeq 0.1$, a suppression factor $n=7$ is enough to have safe predictions for the FV transitions $\ell \to \ell' \gamma$. 
That can be achieved by considering an extra $Z_8$ symmetry under which the flavon charges are ${\cal Q}\left(\phi_e^i\right)=7$ and ${\cal Q}\left(\phi_\nu^i\right)=1$.
The largest possible correction contributing non-trivially to the off-diagonal elements of the Yukawa matrix in Eq.~\eqref{eq:YeA5CP} comes from the replacement $\phi_e^i \to 7\, \phi_\nu^i$ in the leading diagram.
Then, setting $\Lambda_f=515~\sqrt{\kappa}$~GeV in Eq.~\eqref{eq:CD}, in order to account for the discrepancy as discussed in section~\ref{subsec:DipoleStruct}, the following branching fractions are obtained:
\beq
    {\rm BR}\left(\mu \to e\gamma\right) ~ \simeq ~ 8 \times 10^{-15},
\eeq
not far below the foreseen sensitivity of the running MEG-II experiment~\cite{Baldini:2018nnn} (remember that our calculations are affected by $\mathcal{O}(1)$ uncertainties). Moreover, this translates into branching ratios for $\mu\to 3e$ and $\mu\to e$ conversion in nuclei at the $\sim 10^{-16}$ level, within the reach of future LFV searches~\cite{Calibbi:2017uvl}. Obviously, this does not  imply that this kind of $A_5\ltimes {\rm CP}$ models will be necessarily tested by upcoming LFV searches, as it can not be excluded that a different choice of the `shaping' symmetry $Z_N$ would suppress the off-diagonal entries even further.\footnote{Whether there is a limit on the amount of suppression that can be possibly achieved in this way is an interesting question that we postpone to future work.}

\subsection{$U(1)$ toy model for electron and muon $(g-2)$}
\label{sec:toy}
A model based on a $U(1)$ flavour symmetry that accommodates the discrepancies in Eqs.~\eqref{eq:damnew} and~\eqref{eq:daeCs} without inducing large LFV effects has been presented in Ref.~\cite{Calibbi:2020emz}.

The model is an explicit realization of the general idea discussed here, although restricted to two generations. In this model, the charged-lepton masses are produced through the interplay of two processes: the tree-level FN diagram and a radiative correction arising from closing two of the flavon lines in the FN transition.
Here, the radiative corrections, which are usually subleading in other models, are comparable to the tree-level diagram. This is due to a hierarchy among the vevs of the flavons entering the FN diagram and some new inert flavons which do not couple directly to the leptons but only to other fields in the scalar potential. If we define $\varepsilon_{\rm FN} = v_{\rm FN}/\Lambda_f$ and $\varepsilon_{\rm RAD} = v_{\rm RAD}/\Lambda_f$ as the vevs of the different flavons, and given that $m_\ell^{\rm RAD}/m_\ell^{\rm FN}\sim 1/(16\pi^2)\, \varepsilon_{\rm RAD}^2/\varepsilon_{\rm FN}^2$, the radiative contribution can be comparable to the tree-level FN if $ \varepsilon_{\rm RAD}^2 \simeq 16\pi^2\, \varepsilon_{\rm FN}^2$ . 
Then, the same diagram producing the radiative correction to the mass generates a contribution to $\Delta a_\ell$ when a photon is radiated by one electrically charged virtual particle: $\Delta a_\ell/m_\ell^{\rm RAD}\sim 2 m_\ell^{\rm exp}/\Lambda_f^2$.
In this way, the lepton masses and $\Delta a_\ell$, which share a common origin, can provide a non-trivial relation between the dipole contributions to the electron and the muon.

The model presented in \cite{Calibbi:2020emz} has $U(1)\times Z_2$  flavour group.
The electron and muon masses arise from the interplay described above, which can have either sign, whilst $\Delta a_\ell$ is proportional to the radiative mass diagram. In this way, we were able to obtain different contributions to
$\Delta a_e$ and 
$\Delta a_\mu$, even  with  opposite signs. Furthermore, the $U(1)$ charges are assigned so that LFV transitions between electrons and muons would be possible through diagrams with $2n$ and $2n+1$ flavon insertions.
However, the addition of the $Z_2$ symmetry completely eliminates any effective vertex between electrons and muons that would contribute to $\mu\to e\gamma$. 

 However, in this toy model, neither the tau lepton, nor the neutrino sector are explicitly defined. A completion of the model, that would require the inclusion of the third generation of leptons together with a neutrino spectrum that correctly generates the PMNS mixing matrix, seems possible, in principle. In this completion a similar mechanism could also be used to forbid $\tau \to e \gamma$, but would not forbid $\tau \to \mu \gamma$, which, on the other hand, is not so constraining.  

Moreover, the radiative origin of leptonic masses makes possible to reproduce the discrepancy in the muon anomalous magnetic moment with a larger mass for most of the flavons and fermionic mediators, that in this toy model have masses $\Lambda_f\lesssim 3$ ~TeV. The possibility of searching these particles at LHC or future colliders was discussed in \cite{Calibbi:2020emz}.

%=================
\section{Conclusions}
\label{subsec:Conclusions}
%=================
Following the recent confirmation in the Muon g-2 experiment at Fermilab of the muon anomalous magnetic moment discrepancy, in this paper we have analysed the consequences of this result in models with low-scale breaking of a flavour symmetry.

We assume here that the observed discrepancy is fully due to a low-scale flavour symmetry. 
%In these models, it would be possible to generate a different contribution to the electron magnetic moment~\cite{Calibbi:2020emz}, as these contributions are not expected to be exactly proportional to the Yukawa couplings. 
In these models, we have a deep relation between the charged-lepton Yukawa and the dipole matrices, as both matrices have exactly identical transformation properties from the point of view of the flavour symmetry. Extracting global common factors, the different elements in both matrices are, one-to-one, of the same order. However, we expect different $\mathcal{O}(1)$ elements in each entry. This is specially dangerous in off-diagonal entries which will contribute to lepton flavour violation processes.
In this scenario, after diagonalising the charged-lepton Yukawa matrix, off-diagonal entries in the dipole matrix are not eliminated. Then, we expect the same mass scale to be responsible for the anomalous magnetic moment and LFV processes, such that the strong experimental limits on LFV would directly constrain the flavour structure of the  matrices. In practice, these constraints would require an almost diagonal charged-lepton mass matrix, as shown in Eq.~\eqref{eq:Ye-bounds}.
We have applied these constraints to several popular flavour symmetries in the leptonic sector.

Discrete symmetries like $\Delta(27)$, that can explain both the quark and lepton sectors, tend to produce non-diagonal charged-lepton matrices, and therefore we have shown they can not be fully responsible of the $(g-2)_\mu$ discrepancy as they would generate too large contributions to the $\mu \to e \gamma$ decay. Even symmetries like $A_4$, that predict diagonal charged-lepton matrices at leading order, are unable to satisfy the very stringent constraints from LFV when subleading contributions are taken into account. A viable symmetry could be $A_5 \ltimes {\rm CP}$ which can reproduce the observed $\theta_{13}$ mixing without appealing to subleading entries in the charged-lepton matrices. Then, it is possible to protect the diagonal charged-lepton matrices to a sufficient degree.

In summary, if low-energy flavour symmetries are responsible for the observed discrepancy in the muon anomalous magnetic moment, we have strong constraints in the structure of the charged lepton Yukawa matrices. Indeed most of the flavour symmetries are unable to satisfy these stringent bounds and only some symmetries that can protect the diagonal charged-lepton matrices to a high degree, like $A_5 \ltimes {\rm CP}$ are able to explain simultaneously $(g-2)_\mu$ and LFV processes. 

Naturally, all these flavour symmetries can still explain the structure of the fermion mass matrices if they are not the only dynamics responsible of the $(g-2)_\mu$ discrepancy, and the scale of symmetry breaking is pushed to a sufficiently high scale.
However, as we said, if the flavour dynamics is responsible for the discrepancy (or some unspecified new physics whose flavour structure is still dictated by our flavour symmetries), we have shown that only very special symmetries can be viable. It would be interesting to test this possibility further. As we have seen  in section~\ref{subsec:DipoleStruct}, the typical scale suppressing the dipole operator should  be quite below 1~TeV in order to address 
the $(g-2)_\mu$ anomaly. In particular, charged mediator field are expected to have mass around that scale. Hence, there are good prospects for the future runs of the LHC to test this framework.
Furthermore, some of these fields will necessarily couple to muons, providing an excellent physics case for a future muon collider. For model-independent strategies to test the $(g-2)_\mu$ anomaly at a muon collider see e.g.~\cite{Buttazzo:2020eyl,Yin:2020afe,Chen:2021rnl}.
In order to assess the testability of this framework, one should however specify the UV completion (see Refs.~\cite{Calibbi:2012yj,Calibbi:2012at} for a general discussion) within some specific models, which would allow to study the spectrum and the collider signatures following production and decays of such particles. We defer this interesting discussion to future studies. 

\medskip
\noindent {\bf Acknowledgments.}
LC~is partially supported by the National Natural Science Foundation of China under the grant No.~12035008. OV was supported  by Spanish and European funds under MICIU Grant FPA2017-84543-P and from the ``Generalitat Valenciana'' grant PROMETEO2017-033. AM is supported by the Estonian Research Council grants PRG356, PRG434, PRG803, MOBTT86, MOBTT5 and by the EU through the European Regional Development Fund CoE program TK133 ``The Dark Side of the Universe". MLLI acknowledges support from the China Postdoctoral Science Foundation No.2020M670475.
%\newpage
%=========================================================================
\bibliographystyle{JHEP} 
\bibliography{biblio}% Produces the bibliography via BibTeX.
%=========================================================================
%
\end{document}